\documentclass{article}
\usepackage {graphicx}
\begin{document}
\baselineskip .75cm 
\begin{titlepage}
\title{\bf Thermodynamic inconsistency in quasiparticle model - a revisit}       
\author{Vishnu M. Bannur  \\
{\it Department of Physics}, \\  
{\it University of Calicut, Kerala-673 635, India.} }   
\maketitle
\begin{abstract}

Widely studied quasiparticle models for quark gluon plasma is revisited here to understand the statistical mechanics and thermodynamics of the system. We investigate the statistical mechanics and thermodynamics inconsistencies involved in these models and their consequences in the observables. Quasiparticle model is a phenomenological model with few parameters and by adjusting them all models fit the results of lattice gauge simulation of gluon plasma \cite{bo.1}. However, after fixing 2 of the 3 parameters of the model by physical arguments, only one quasiparticle model, which is consistent with both statistical mechanics and thermodynamics, fits the Bielefeld lattice data \cite{bo.1}. The same model also fits the recent lattice results of Wuppertal-Budapest group \cite{fo.1}, which deals with precision SU(3) thermodynamics for a large temperature range, reasonably well.       
\end{abstract}
\vspace{1cm}
                                                                                
\noindent
{\bf PACS Nos :} 12.38.Mh, 05.30Jp, 12.39.Mk, 12.38.Gc, 05.70.Ce \\
{\bf Keywords :} Equation of state, gluon plasma, quasiparticle model. 
\end{titlepage}
\section{Introduction :}

Quark gluon plasma (QGP) is a non-ideal system and hence the formulation of statistical mechanics (SM)and thermodynamics (TD) is a challenging one. SM and TD of ideal system is straight forward, but the equilibrium studies of non-ideal system is still underdeveloped because of the difficulties due to mutual interaction among the constituents. Of course, there are theories like classical and quantum Mayer cluster expansion \cite{pa.1}, but they are not fully successful for QGP because of the strong interaction \cite{ba.1}. Mayer's cluster expansion is valid for weakly coupled system, whereas QGP seems to be strongly coupled \cite{ba.2}. Therefore, one goes for phenomenological models like strongly coupled quark gluon plasma \cite{ba.2,bt.1},  quasi-particle models \cite{gs.1,pe.1,go.1,pe.2,ba.3,bl.1,br.1,ga.1,sr.1,yi.1,zh.1,co.1,zu.1,ji.1}
, etc. with few fitting parameters. Model results are fitted with the results of the simulation of lattice gauge theory (LGT) for QGP \cite{bo.1, pn.1, fo.1}. Here we study and compare different quasiparticle models and discuss SM and TD consistencies of each models. 

Next section we discuss the quasiparticle model which was first introduced. The modification of this model by eliminating the TD inconsistency is presented in Section III. Another alternative way without TD and SM inconsistencies is presented in Section IV. Discussions and conclusions is presented in Section V. \\
\section{Quasiparticle Model - I (QPM-I):} 

Quasi-particle model for QGP was first introduced by Goloviznin and Satz \cite{gs.1}, and then Peshier {\it et. al.} for gluon plasma \cite{pe.1}. Following the recent article by Castorina, Miller and Satz \cite{ca.1, bu.1}, one starts from the expression for pressure, \\
\[ P = - T \,d_f \int_0^{\infty} dp \,p^2 \,\ln(1 - \exp(-\frac{\sqrt{p^2 + m^2}}{T})) \,\, , \]
\begin{equation}  
  = d_f \,T^4 \,\sum_{l=1}^{\infty} (\frac{m l}{T})^2  \,\frac{K_2 (\frac{m l}{T})}{l^4} \,\,, 
\end{equation}
where $d_f \equiv \frac{8}{\pi^2}$ and T is the temperature. $K_2 (x)$ is the modified Bessel function of order 2. $m$ is the thermal mass which is a function of temperature. Basic idea behind the quasiparticle model is that the effects of mutual interaction among the constituents in nonideal plasma is taken in the form of effective mass or thermal mass and treat the system as ideal. There are different phenomenological models for thermal masses and here we use the recent model, \cite{ca.1}, \\
\begin{equation}
m(T) = \frac{a}{(t-1)^c} + b\,t \,\, ,
\end{equation}
where $t \equiv \frac{T}{T_c}$ and $a, b, c$ are parameters of the model. $T_c$ is the phase-transition temperature to gluon plasma.  The structure of this $m(T)$ is discussed, in detail, in Ref. \cite{ca.1} which involves the main properties of phase transition near critical temperature and proper high temperature limit for large temperature. 
 
All other thermodynamic quantities like energy density, speed of sound, etc. are derived from pressure using thermodynamic relations. For e.g., the energy density, $\varepsilon$, is \\
\begin{equation}
\varepsilon = d_f \, T^4 \,\sum_{l=1}^{\infty} \frac{1}{l^4} \left[ 3 (\frac{m l}{T})^2 \, K_2 (\frac{m l}{T}) + l\, (\frac{m}{T} - \frac{dm}{dT}) \,(\frac{m l}{T})^2 \, K_1 (\frac{m l}{T}) \right] \,\,.
\end{equation}
Let us call this model QPM-I. 

However, above expression for energy density leads to statistical mechanics inconsistency. We know that when we treat an equilibrium system as canonical ensemble (CE) or grand canonical ensemble (GCE) the energy of the system is not constant and fluctuates and we define average energy \cite{pa.1} and hence energy density as \\
\begin{equation}
\varepsilon_{SM} = d_f \, T^4 \,\sum_{l=1}^{\infty} \frac{1}{l^4} \left[ 3 (\frac{m l}{T})^2 \, K_2 (\frac{m l}{T}) + \, (\frac{m l}{T})^3 \, K_1 (\frac{m l}{T}) \right] \,\,.
\end{equation}
Therefore, we see that $\varepsilon$ $\ne$ $\varepsilon_{SM}$ and hence SM inconsistency. This model violates the standard SM of CE or GCE system. 

But, since $m(T)$ has few parameters, we can tune them to fit LGT data for pressure and them obtain other TD quantities, using TD relations. Hence it is TD consistent. Thus we see that this model is TD consistent, but SM inconsistent. 
\begin{figure}[h]
\centering
\includegraphics[height=8cm,width=12cm]{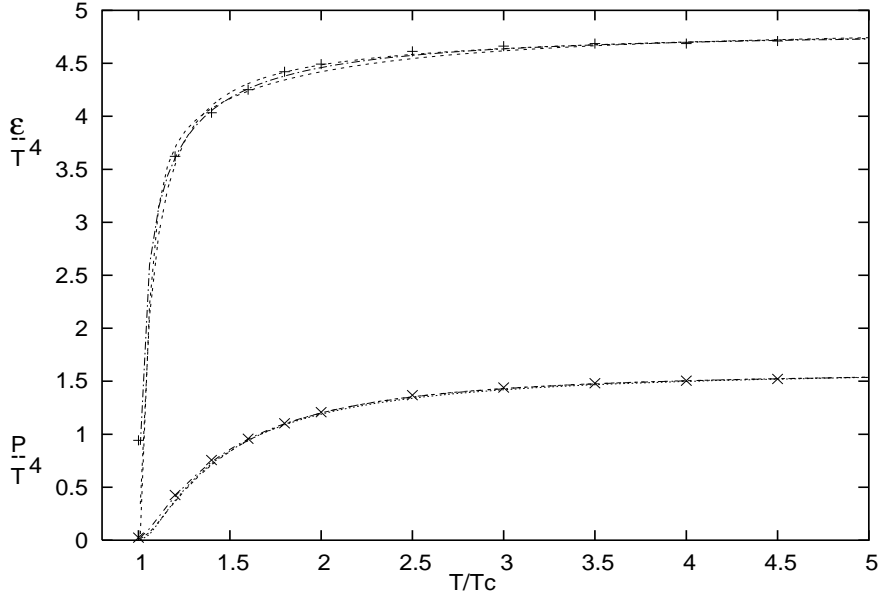}
\caption{ Plot of $\varepsilon/T^4$ and $P/T^4$ as a function of $T/T_c$ for gluon plasma from 3 quasiparticle models QPM-I, QPM-II, QPM-III and compared with LGT results of Ref. \cite{bo.1} (symbols).}  
\end{figure}
\section{Quasiparticle Model - II (QPM-II):} 

In this model, QPM-II, we follow Peshier {\it et. al.} \cite{pe.1} where one starts from pressure, just like the earlier model, \\
\begin{equation}
  P = d_f \,T^4 \,\sum_{l=1}^{\infty} (\frac{m l}{T})^2  \,\frac{K_2 (\frac{m l}{T})}{l^4} \,\,.
\end{equation}
However, the energy density $\varepsilon$ is \\
\begin{equation}
\varepsilon = d_f \, T^4 \,\sum_{l=1}^{\infty} \frac{1}{l^4} \left[ 3 (\frac{m l}{T})^2 \, K_2 (\frac{m l}{T}) + \, (\frac{m l}{T})^3 \, K_1 (\frac{m l}{T}) \right]  = \varepsilon_{SM} \,\,, 
\end{equation}
which is different from QPM-I. Hence it is SM consistent. But the drawback is that it is TD inconsistent as pointed out by various authors \cite{go.1,pe.2,ba.3}.  

To solve this problem, borrowing idea from the bag model of hadron spectroscopy, one introduces bag pressure term $B(T)$ in the expression for pressure as, \\
\begin{equation}  
 P  = d_f \,T^4 \,\sum_{l=1}^{\infty} (\frac{m l}{T})^2  \,\frac{K_2 (\frac{m l}{T})}{l^4} - B(T)\,\,, \label{eq:p}
\end{equation}
and energy density, \\
\begin{equation}
\varepsilon = d_f \, T^4 \,\sum_{l=1}^{\infty} \frac{1}{l^4} \left[ 3 (\frac{m l}{T})^2 \, K_2 (\frac{m l}{T}) + \, (\frac{m l}{T})^3 \, K_1 (\frac{m l}{T}) \right] + B(T) \,\,. \label{eq:e}  
\end{equation}
This $B(T)$ was interpreted as temperature dependent bag pressure, vacuum pressure, residual interaction energy, etc. Then one exploit $B(T)$ to solve the TD inconsistency problem by demanding that pressure and energy density, Eqs. (\ref{eq:p}) and (\ref{eq:e}), be TD consistent. This leads to TD consistent relation, \\
\begin{equation}
\frac{dB}{dT} = - m\,\frac{dm}{dT} \, d_f \, T^2 \,\sum_{l=1}^{\infty} \frac{1}{l^2} (\frac{m l}{T}) \, K_1 (\frac{m l}{T}) \,\,, \label{eq:tdc} 
\end{equation}
and hence $B(T)$ is related to $m(T)$. Thus the QPM-II is now TD and SM consistent. 
\begin{figure}[h]
\centering
\includegraphics[height=8cm,width=12cm]{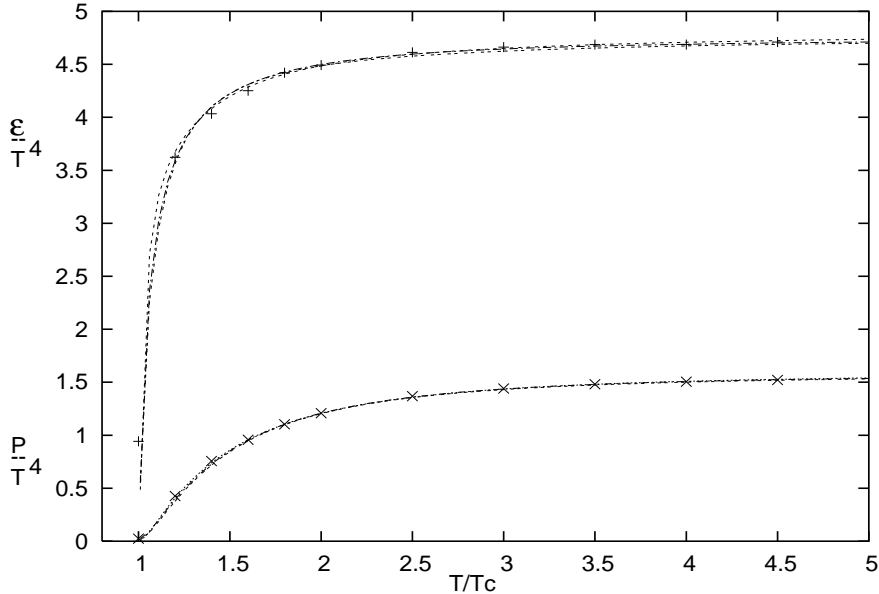}
\caption{ Plot of $\varepsilon/T^4$ and $P/T^4$ as a function of $T/T_c$ for gluon plasma from 3 quasiparticle models QPM-I, QPM-II, QPM-III at Boltzmann limit and compared with LGT results of Ref. \cite{bo.1} (symbols).}  
\end{figure}
\section{Quasiparticle Model - III (QPM-III):}

Here we follow yet another approach, QPM-III, which follow straight forward from standard SM. Once the interaction effects of nonideal system is absorbed in thermal mass and treat the resultant system as ideal system in CE or GCE. We know that in CE or GCE, one defines average energy \cite{pa.1} and hence we start from energy density, \\
\begin{equation}
\varepsilon = d_f \, T^4 \,\sum_{l=1}^{\infty} \frac{1}{l^4} \left[ 3 (\frac{m l}{T})^2 \, K_2 (\frac{m l}{T}) + \, (\frac{m l}{T})^3 \, K_1 (\frac{m l}{T}) \right] = \varepsilon_{SM} \,\,, 
\end{equation}
and hence SM consistent. Pressure is obtained from the TD relation, \\
\begin{equation}
\varepsilon = T \frac{dP}{dT} - P \,\,,
\end{equation}
on integration and thus it is TD consistent also. Therefore, this model is both SM and TD consistent and no need of $B(T)$. Of course, one can include $B(T)$ in this model also, but then $B(T)$ is free function of $T$. As we have shown in Ref. \cite{ba.4} this model gives QPM-II with the help of an extra constraint equation, Eq. (\ref{eq:tdc}). 
\begin{center}
\begin{tabular}{||c|c|c|c||c|c|c||}
\hline 
&\multicolumn{3}{|c}{Exact} &\multicolumn{3}{|c||}{Boltzmann limit} \\
\hline
Model &$a/T_c$ &$b/T_c$ &$c$ &$a/T_c$ &$b/T_c$ &$c$ \\  
\hline
QPM-I &1.6 &0.53 &.45 &2.0 &.22 &.35 \\  
\hline
QPM-II &0.95 &0.75 &.385 &1.3 &0.4 &.3  \\
\hline
QPM-III &0.65 &0.88 &0.385 &0.85 &0.5 &0.385 \\  
\hline
\end{tabular} \\[.5cm]
{\bf Table 1} \\[.5cm]   
\end{center}

\section{Discussions and conclusions:} 

Thus we saw that QPM-I is SM inconsistent, but by exploiting the phenomenological parameters we can fit the LGT data \cite{bo.1} of gluon plasma for pressure. Values of parameters are tabulated in Table-I. But it is interesting that even after the violation of SM, by adjusting three parameters one can get reasonably good fit as shown in Fig. 1. Using the 3 fitting parameters other TD quantities like $\varepsilon$ is evaluated from the TD relations and also plotted in Fig. 1 and compared with LGT data and fits well as it should. 

Next we discuss the results of QPM-II. It is SM consistent and also TD consistent by TD consistent relation, and again reasonably good fit to LGT data for pressure and energy density are obtained by adjusting 3 parameters which are tabulated in Table-I. Plots are given in Fig. 1. 

Results of QPM-III, which is both SM and TD consistent, is also plotted in Fig. 1, by fitting energy density with LGT data by adjusting 3 parameters. Values of fitted parameters are given in Table -I. Again reasonably good fits are obtained. 

It is interesting to note that even the single term in the summation for the expression of thermodynamic quantities, pressure or energy density, in all three models, also fits the LGT data by different sets of fitting parameters given in Table-I (right). These results are equivalent to the Boltzmann limit discussed in Ref. \cite{ca.1}. Plots are compared in Fig. 2.  
\begin{figure}[h]
\centering
\includegraphics[height=8cm,width=12cm]{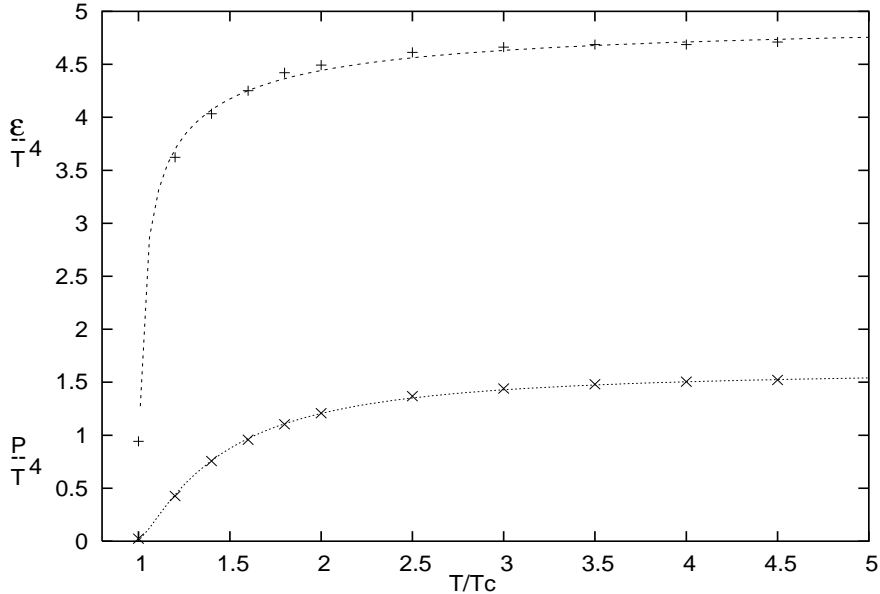}
\caption{ Plot of $\varepsilon/T^4$ and $P/T^4$ as a function of $T/T_c$ for gluon plasma from our model QPM-III with running $b(t)$ and compared with LGT results of Ref. \cite{bo.1} (symbols).}  
\end{figure}

In the Ref.\cite{ca.1} the reason for the 3 parameters in $m(T)$ is discussed. $c$ is related to the critical exponent of phase transition and is estimated to be $0.41$ for SU(2). Parameter $b$ is introduced based on the fact that thermal mass is supposed to be proportional to $T$ as $T \rightarrow \infty$. We may even demand that at high temperature $m(T)$ may goes as $\sqrt{\frac{3}{2}} \omega_p$  where $\omega_p$ is the plasma frequency which follows from perturbative results of quantumchromodynamics (QCD). Therefore, we may take  
\begin{equation}
m(T) = \frac{a}{(t-1)^c} + b(t) \,t \,\, ,
\end{equation}
where $b(t)/T_c = \sqrt{2 \pi \alpha_s (t)}$, which is not constant as in Ref. \cite{ca.1}, but running constant because of $\alpha_s(T)$. We used the QCD perturbative result, $\omega_p^2 = 4 \pi \alpha_s (T) T^2 /3$. We may take 2-loop perturbative coupling constant,
\begin{equation} \alpha_s (T) \equiv \frac{g^2}{4 \pi} = \frac{6 \pi}{(33-2 n_f) \ln (T/\Lambda_T)}  
\left( 1 - \frac{3 (153 - 19 n_f)}{(33 - 2 n_f)^2} 
\frac{\ln (2 \ln (T/\Lambda_T))}{\ln (T/\Lambda_T)} \right)  \;, \label{eq:as} 
\end{equation}
where $\Lambda_T = 0.14 \,T_c$ from the lattice studies Ref. \cite{bo.1} with which we compare our results.  Assuming that SU(3) also have the similar phase transition as SU(2), we may take $c=0.41$ for gluon plasma. Now we have only one free parameter, $a$, and by adjusting it to fit to LGT data using our model QPM-III, we again get good fit as plotted in Fig. 3 for $a = 0.45$. Advantage here is that 2 of 3 parameters of $m(T)$ is fixed by physical arguments. At present we don't know to estimate the value of $a$. We also tried to fit LGT data using QPM-I and QPM-II with the same $\Lambda_T = 0.14$ and $c=0.41$, but by adjusting $a$. However, both of these models fail to get good fit to LGT data.  
\begin{figure}[h]
\centering
\includegraphics[height=8cm,width=12cm]{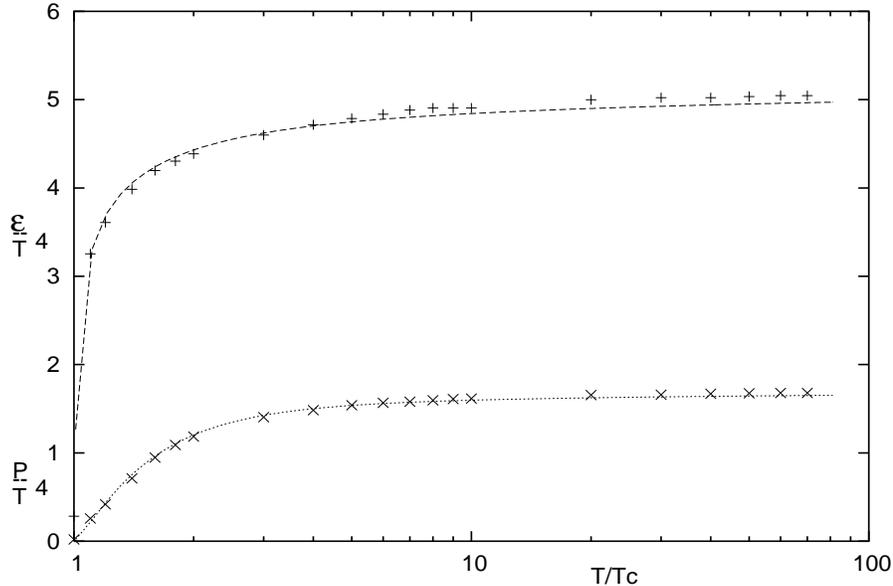}
\caption{ Plot of $\varepsilon/T^4$ and $P/T^4$ as a function of $T/T_c$ for gluon plasma from our model QPM-III with running $b(t)$ and compared with LGT results of Ref. \cite{fo.1} (symbols).}  
\end{figure}

It is further interesting to note that the same model, QPM-III, also fits the recent lattice results of Wuppertal-Budapest group \cite{fo.1}, which deals with precision SU(3) thermodynamics for a large temperature range, reasonably well as shown in Fig. 4. Here also $a=0.45$ and $c=0.41$, but $\alpha_s(T)$ is the 3-loop running coupling constant used in the same reference \cite{fo.1}. Our result is less than 2 \% lower than that of lattice for large temperature as can be seen from Fig. 4. There is no much difference even if we use earlier $\alpha_s(T)$, Eq.(\ref{eq:as}), with the same $\Lambda_T = 0.14 \,T_c$.  
              
In conclusion, all three quasiparticle models \cite{ca.1, pe.2, ba.3}, considered here, fit the LGT data on gluon plasma \cite{bo.1} reasonably well by adjusting three parameters of a phenomenological thermal mass proposed recently by Castorina, Miller and Satz \cite{ca.1}. However, only one model \cite{ba.3}, QPM-III,  which is both statistical mechanics and thermodynamics consistent, fits the LGT data after fixing 2 of 3 parameters by physical arguments and adjusting the remaining one free parameter. It is remarkable that the same model, QPM-III, which fits Bielefeld lattice data \cite{bo.1}, also gives good fit to the recent lattice results of Wuppertal-Budapest group \cite{fo.1} which includes data for larger range of temperature ($>5 T_c$).        

\end{document}